\def\bJ{{\beta_{_{J}}}}
\def\TsCO{T^{\star}_{co}}

\def\iSW{i_{SW}}

\documentclass[aps,twocolumn,superscriptaddress,showkeys]{revtex4-1}
\usepackage{multirow}

\usepackage{epsfig}
\usepackage{natbib}
\usepackage{amsmath}
\usepackage{times}
\usepackage{psfrag}
\usepackage{epstopdf}
\usepackage{braket}

\usepackage{graphicx}
\usepackage{subfigure}
\usepackage{float}
\newcommand{\hg}{ }

\begin{document}

\title{Anomalous transport effects on switching currents of graphene-based Josephson junctions}

\author{Claudio Guarcello }
\affiliation{ SPIN-CNR, Via Dodecaneso 33, 16146 Genova, Italy}
\affiliation{ NEST, Istituto Nanoscienze-CNR and Scuola Normale Superiore, Piazza S. Silvestro 12, I-56127 Pisa, Italy}
\affiliation{ Radiophysics Dept., Lobachevsky State University, Nizhniy Novgorod, Russia}

\author{Davide Valenti}
\affiliation{Dipartimento di Fisica e Chimica, Interdisciplinary Theoretical Physics Group, Universit\`a di Palermo and CNISM, Unit\`a di Palermo, Palermo, Italy}

\author{Bernardo Spagnolo }
\affiliation{Dipartimento di Fisica e Chimica, Interdisciplinary Theoretical Physics Group, Universit\`a di Palermo and CNISM, Unit\`a di Palermo, Palermo, Italy}
\affiliation{ Radiophysics Dept., Lobachevsky State University, Nizhniy Novgorod, Russia}
\affiliation{Istituto Nazionale di Fisica Nucleare, Sezione di Catania, Catania, Italy}

\author{Vincenzo Pierro}
\affiliation{Dipartimento di Ingegneria, Universit\`a del Sannio, Benevento, Italy}

\author{Giovanni Filatrella}
\affiliation{Dipartimento di Scienze e Tecnologie and CNISM, Unit\`a di Salerno, Universit\`a   del Sannio, Benevento, Italy}

\begin{abstract}
We explore the effect of noise on the ballistic graphene-based small Josephson junctions in the framework of the resistively and capacitively shunted model.
We use the non-sinusoidal current-phase relation specific for graphene layers partially covered by superconducting electrodes. 
The noise induced escapes from the metastable states, when the external bias current is ramped, give the switching current distribution, i.e. the probability distribution of the passages to finite voltage from the superconducting state as a function of the bias current, that is the information more promptly available in the experiments. 
We consider a noise source that is a mixture of two different types of processes: a Gaussian contribution to simulate an uncorrelated ordinary thermal bath, and non-Gaussian, $\alpha$-stable (or L\'evy) term, generally associated to non-equilibrium transport phenomena.
We find that the analysis of the switching current distribution makes it possible to efficiently detect a non-Gaussian noise component in a Gaussian background. 

\end{abstract}

%
\maketitle

\section{Introduction}
\label{Intro}

The Josephson Junction (JJ) electrical behavior is governed by a quantum variable, the gauge invariant phase difference between the macroscopic phases of the two superconductors forming the junction.
Its dynamics, as described by the celebrated Josephson equations~\cite{Jos62,Jos74}, is not directly accessible, and only indirect electrical measurements (essentially, the current and the voltage) can be actually monitored.
In particular it is possible to retrieve the current at which the passage from the superconducting state to the finite voltage occurs in JJ, usually called switching current (SC). 
Repeating the measurements in the presence of a random disturbance, or because of quantum effects~\cite{Pierro16}, the junction can switch to the finite voltage at slightly different current levels, thus producing a distribution of SCs.
Conversely, the analysis of the distribution of the switching currents can be used to reveal the presence of noise or quantum effects. 
In fact, the analysis of the SCs cumulants have been employed to detect and quantify noise~\cite{Ank05,Pel07,Tim07}, or to ascertain its features~\cite{Gol10}.
In this context, it is attractive the idea to use a JJ as a device that is capable to discern the non-Gaussian, in particular Poissonian, character of the non intrinsic noise~\cite{Pek04}.

We propose to extend these ideas to graphene-based JJs affected by L\'evy noise, that is non-Gaussian fluctuations characterized by the so-called L\'evy flights~\cite{Dub07,Dub08,Dub09}, exploiting the information content of the SC distributions, in analogy to the case of driven JJs~\cite{Fil10,Add12}.
There are several reasons to do so. 
From a general point of view, stable non-Gaussian noise sources are interesting exceptions to the general rule that noise is characterized by finite variance.
It is therefore valuable to have a tool that can accurately discriminate the presence of small amounts of "fat tails" disturbances to get an estimate of their contribution in the noise background.
Moreover, JJs as detectors of non-Gaussian sources offer the advantage that, being superconducting devices, can produce little thermal noise, for the intrinsic resistance of the junction is low and the temperature can be decreased as much as it is necessary or possible. 
JJs are therefore well suited to be used as on chip detector of current fluctuations to characterize a noise source that feeds the junction.
The interest also resides in material issues, for it has been noticed~\cite{Cos12} that graphene-based JJs can exhibit anomalous SCs with premature switches.
Early escapes suggest the presence of anomalous disturbances, that are likely to be unrelated to thermal fluctuations.
Recently, it has also been proposed that the particular electron-electron interaction of the graphene electronic can produce a peculiar response to a laser source, namely, a L\'evy flights distribution~\cite{Bri14}. 
Yet, a graphene stripe with anisotropically distributed on-site impurities is demonstrated to reveal L\'evy flight transport in the stripe direction~\cite{Gat16}.
Therefore, a graphene based device in this configuration could be at the same time both the source of the noise and the detector to reveal its presence.
{\hg Shortly, this work has been prompted by the observation that L\'evy noise has been postulated only in the specific case of graphene-based JJ \cite{Cos12,Bri14,Gat16}, and that to reveal its presence might be of particular relevance for material issues that pertain graphene.}

Determining the properties of the L\'evy noise from the promptly available data, the SC distributions, offers a qualitative advantage: the data can be indeed collected in a pre-established time, because the bias ramp gives a maximum time for each point, $T = 1/v_b$.
By contrast, according to the nature of the L\'evy flights, the extreme fluctuations that cause the flat tails of the L\'evy noise distributions are rare but essential to characterize the tails; therefore,  exceedingly long expectation times are usually required to build up a robust statistical analysis.
The special bias scheme to record SC, that guarantees that the data are collected in a maximum finite time $1/v_b$, that is determined by the experimental set-up is shown in the inset of Fig. \ref{Figschematic}a.

To implement JJ as noise detectors, however, requires a careful analysis \cite{Gra08,Aug10,Gua13,Val14,Gua15,Spa15,Gua16}.
L\'evy flights sources are expected to induce a distortion of the cumulants of the SC distribution (with respect to the Gaussian case)~\cite{Gra08}. 
Here we propose to go beyond the analysis of the cumulants, to retrieve the properties of the noise exploiting the full information content of the SCs distribution.
The direct analysis is more convenient, especially to quantify the amplitude of the L\'evy noise.

The work is organized as follows. In Sect.~\ref{Model} we set the stage for the analysis, discussing the basic equations that govern the system.
In Sect.~\ref{sec:SCanalysis} we discuss the methods to analyze the SC resulting from the noisy system and we collect the results of simulations.
Sect.~\ref{Conclusions} concludes.


\section{The Model}

\label{Model}

We investigate a dissipative, current biased short JJ, within the resistively and capacitively shunted junction (RCSJ) model.
For the normalized current, the basic equation reads~\cite{Bar82}
\begin{equation}
\frac{d^2 \varphi (t)}{dt^2}+\bJ \frac{d \varphi (t)}{dt} + i_{\varphi}(t) = i_{f}(t) + i_b(t ).
\label{RCSJ}
\end{equation}
Here, the currents $i$ are normalized to the critical current of the contact $I_c$, $\beta_J$ is the standardized friction parameter, time is normalized to the inverse of the plasma frequency $\omega_{p_0}^{-1} = ( \hbar C/2eI_c)^{1/2}$, where $C$ its capacitance, $\hbar$ is the reduced Planck constant, and $e$ is the electron charge. 

This governing equation consists of three parts: i) the Josephson elements at the left hand side, namely, the capacitive term, the dissipative contribution, and the Josephson supercurrent $i_{\varphi}$, ii) a noise source $i_f(t)$, and iii) the external current bias $i_b(t)$. 
In the following we describe these components.

\subsection{The graphene based Josephson elements}
\label{sec:JJcurrent}

\begin{figure}{ht}
\centering \hspace{1.5cm}
(a) \\
\includegraphics[width=0.47\textwidth]{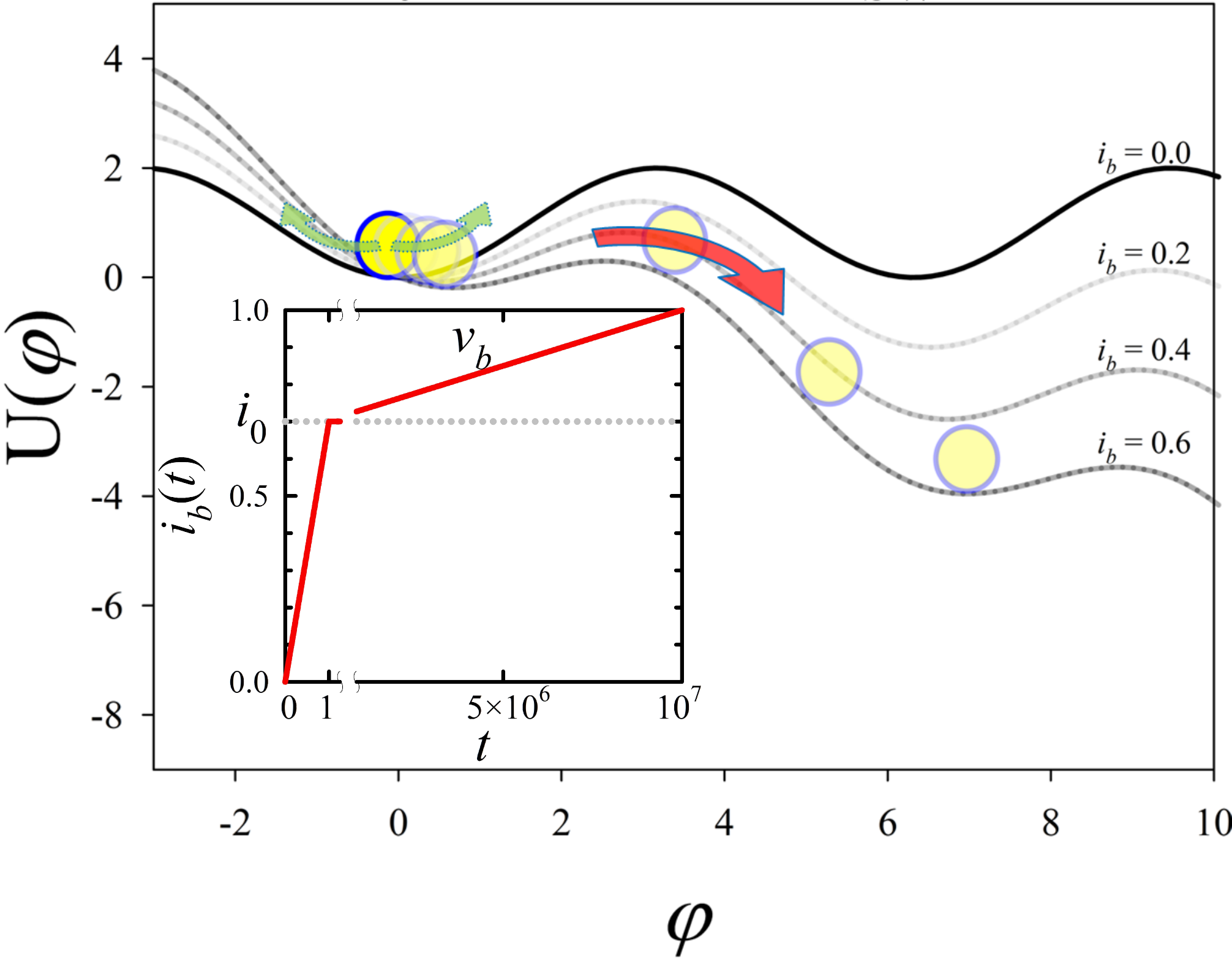} \hspace{0.7cm}\\
 (b) \\
\includegraphics[width=0.35\textwidth]{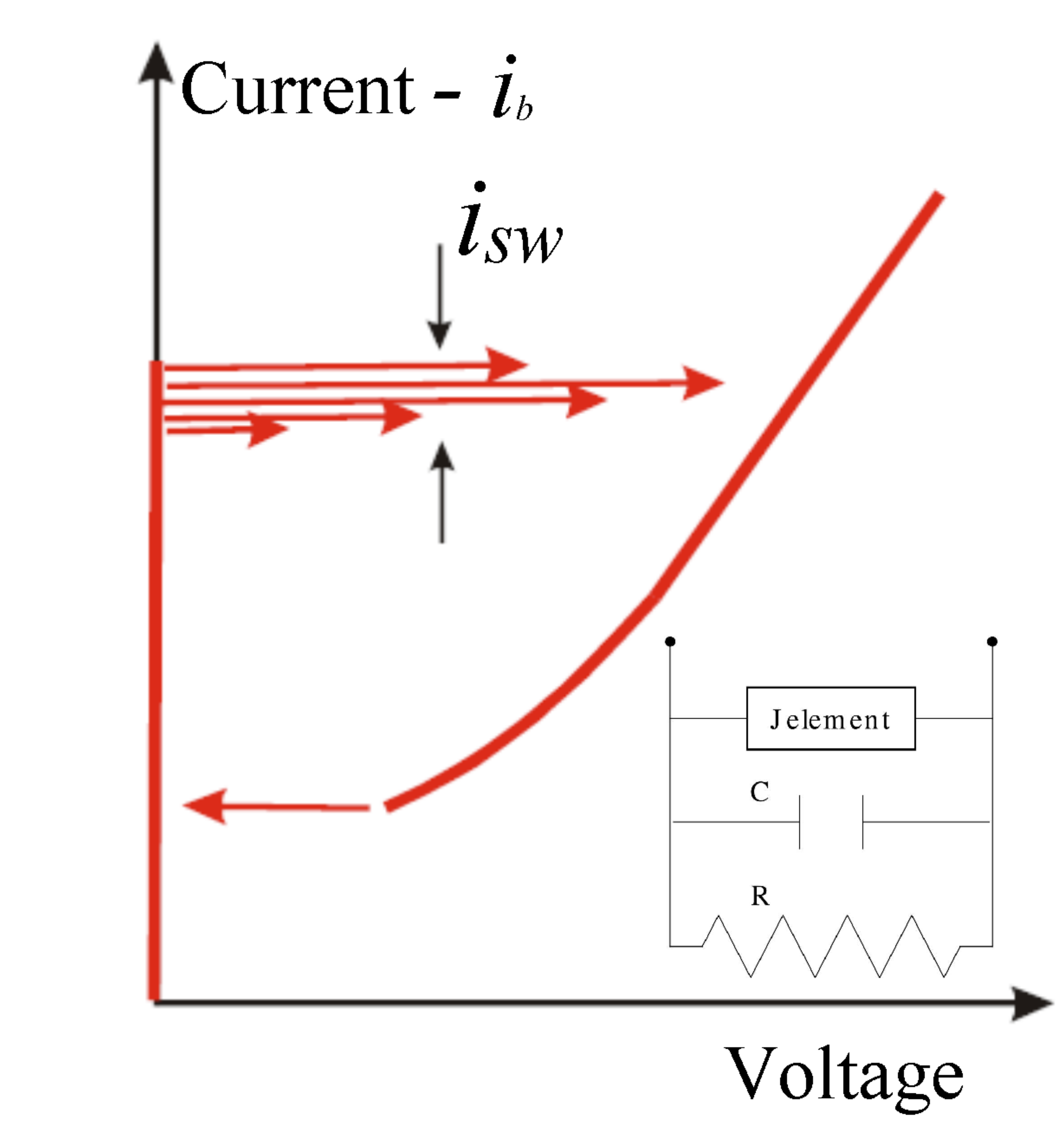}
\caption{(Color online) (a) Mechanical analog of a JJ, corresponding to the representative point of the system running down along the washboard potential as the bias current is increased. Due to noise induced fluctuations, the overcoming of the potential barrier occurs in correspondence of a bias current $i_b<1$. In inset: schematic of the bias current sweeping method. The current is fast swept until the initial bias $i_0$, and then slowly swept (at a rate $v_b$) up to the critical value $i_b=1$. 
{\hg The time scale of the inset is normalized to the inverse of the plasma angular velocity $\omega_{p_{0}}$, see Eq.(\ref{graphenePlasmaFreq}). 
Notice the broken axis, that is necessary to visualize the initial fast ramp. }
(b) I-V scheme: the bias (vertical axis) is ramped, and it is recorded the current at which a passage to the finite voltage occurs. The switching current corresponds to the passage over the maximum of the potential showed in the panel a. In inset: the electrical equivalent of a JJ.}
\label{Figschematic}
\end{figure}

In a traditional tunnel junction the Josephson normalized current contribution of Eq.~(\ref{RCSJ}), $i_\varphi$, is governed by the non linear Josephson element
\begin{equation}\label{NormalIc}
i_{\varphi}(t)=\frac{I(\varphi)}{I_c}=\sin \left [ \varphi\left ( t \right ) \right ]. 
\end{equation}

A short ballistic graphene-based JJ exhibits a more complicated current-phase relation, that, in the low temperature regime, i.e., $T\lesssim T_c/4$, $T_c$ being the critical temperature of the junction, reads~\cite{Bee06, Hag10,Gua15}
\begin{equation}
 i_{\varphi}(t)=\frac{I(\varphi)}{I_c}=\frac{2}{1.33}\cos \left ( \frac{\varphi }{2} \right )\tanh^{-1}\left [ \sin \left (\frac{\varphi }{2} \right) \right ].
 \label{grapheneIphi}
\end{equation}

The critical current $I_c$ is connected with the junction parameters, namely, the superconductive excitation gap $\Delta_0$ and the length and width of the junction $L$ and $W$, respectively, through~\cite{Bee06}
\begin{eqnarray}
I_c&=&1.33 \frac{e\Delta_0}{\hbar}\frac{W}{\pi L}.
\label{grapheneIc}
\end{eqnarray}

It is convenient to introduce the mechanical analogous, in which a graphene-based JJ is assimilated to a representative point on a potential derived from the peculiar current-phase relation (\ref{grapheneIphi})~\cite{Lam11,Gua15}
\begin{eqnarray}
 \label{grapheneWB}
\nonumber U(\varphi,t ) \textup{=} &-&E_{J_0} \Biggr\{ -\frac{2}{1.33} \times \nonumber \\
 &\times & \biggr\{ 2\sin\left (\frac{\varphi (t) }{2} \right ) \tanh^{-1} \left [ \sin\left (\frac{\varphi (t) }{2} \right ) \right ] +\nonumber \\
&+&\ln \left [ 1- \sin^2\left (\frac{\varphi (t) }{2}\right )\right] \biggr\} +i_b(t)\varphi (t) \Biggr\},
\end{eqnarray}
$E_{J_0}=\hbar I_c/2e$ being the Josephson energy.
{\hg It is the potential (\ref{grapheneWB}) that characterizes the graphene-based JJ and that specifically distinguishes the JJ with respect to the traditional washboard potential of traditional JJ \cite{Bar82}}

The small oscillations frequency reads~\cite{Lam11,Gua15}
\begin{equation}
\nonumber \omega_p = \frac{\omega_{p_{0}}}{\sqrt{1.33}}\sqrt{1-\sin\left ( \frac{\varphi_{min}}{2} \right )\tanh ^{-1}\left [ \sin\left ( \frac{\varphi_{min}}{2} \right ) \right ]},
\label{graphenePlasmaFreq}
\end{equation}
where $\varphi_{min}$ is the phase value at a minimum of $U ( \varphi ,t )$.
In the mechanical interpretation, $\omega_p$ is the natural frequency of the system that in the Josephson jargon is named plasma frequency. 

In the formulation of the potential (\ref{grapheneWB}) it is assumed that the bias current $i_b(t)$ depends upon the time, but its dependence is slow with respect to the JJ timescale $\omega_{p_0}^{-1}$.
Next Sect.~\ref{sec:noise} is devoted to the effect of noise in this adiabatic approximation, while the effect of the bias current will be discussed in Sect.~\ref{sec:current}.

The circuit model behind Eq.~(\ref{RCSJ}), see the inset of Fig.~\ref{Figschematic}b, consists of a Josephson tunnel current in parallel with a capacitor and a resistance.
The coefficients of the linear elements of Eq.~(\ref{RCSJ}) are normalized so that the capacitance reads $1$, and the resistive term is governed by the damping parameter $\bJ $
\begin{equation}
\bJ =\frac{1}{\omega_{p_0}R_NC} = \beta_c^{-1/2},
\label{alpha}
\end{equation}
\noindent 
where $R_N$ is the normal resistance  and $\beta_c$ is the McCumber parameter~\cite{Bar82}. 
The dissipation in the nonlinear system slightly alters the proper resonance in Eq.~(\ref{graphenePlasmaFreq})

\begin{equation}
\omega_{res}=\omega_p\left \{ \sqrt{1+\bJ ^2}-\bJ \right\}.
\label{omegaR}
\end{equation}
\noindent

\subsection{The noise source} 
\label{sec:noise}

The term $i_f(t)=I_f(t)/I_c$ in Eq.~(\ref{RCSJ}) represents a random contribution due to the current noise, normalized to the critical current $I_c$. 
For the L\'evy noise the stochastic model is obtained with the approximated finite independent increments~\cite{Chechkin00}.
The random current is modeled as a mixture of a standard Gaussian white noise and a L\'evy process.
This contribution has been proposed to model anomalies in the transport properties, as those described in Ref.~\cite{Bri14}.
{\hg Thus, even if the  methods described in this work could be in principle applicable to generic JJs, the special form of the noise  that we investigate has been, so far, only postulated in graphene-based JJs.}

When both Gaussian and L\'evy flights fluctuations are considered, the stochastic independent increments read
\begin{equation}
\Delta i_f \simeq 
 \left( \gamma_G \Delta t \right) ^{1/2} N\left(0,1 \right) + 
 \left( \gamma_L \Delta t \right) ^{1/\alpha}S_{\alpha}\left(1,0,0\right).
\label{GLFincr}
\end{equation}
Here the symbol $N\left(0, 1 \right)$ denotes a normal random variable with zero mean and unit standard deviation.
Furthermore, $S_{\alpha}(1, 0, 0)$ denotes a \emph{standard} $\alpha$-stable random L\'evy variable~\cite{Gua16}. These distributions are symmetric around zero, and for $\alpha<2$ are characterized by an asymptotic long-tail power law behavior with exponent $-(1+\alpha)$, while the limit $\alpha=2$ case corresponds to the Gaussian distribution. Moreover, the $\alpha=1$ case corresponds to the well-known Cauchy-Lorentz distribution.
The algorithm used to simulate L\'evy noise sources is that proposed by Weron~\cite{Wer96} for the implementation of the Chambers method~\cite{Cha76}. The stochastic dynamics of the system is analyzed by integration of Eq.~(\ref{RCSJ}) with a finite-difference method. 
The time step is fixed at $\Delta t = 10^{-2}$, the maximum integration time is $t_{max} = 10^7$, and the number of numerical realization to produce the SC distributions is $N=10^4$.

Let us give some physical considerations on the parameter $\gamma_G$ of Eq.~(\ref{GLFincr}). 
If we consider the pure white noise case $\gamma_L=0$, the statistical properties of the current fluctuations, in physical units are determined by
\begin{equation}
E\left [ I_f\left ( \tilde{t}\right ) \right ] = 0 \nonumber \\
\end{equation}
and
\begin{equation}
E\left [ I_f\left (\tilde{t}\right)I_f\left (\tilde{t}+\tilde{\theta}\right) \right ] =
2\frac{kT}{R_N}\delta \left (\tilde{\theta} \right ), 
\label{WNProperties}
\end{equation}
where $T$ is the temperature of the system, $k$ is the Boltzmann constant, $E[\cdot ]$ is the expectation operator, $\tilde{t}$ and $\tilde{\theta}$ denote physical times. 
In our normalized units, the same equations become 
\begin{eqnarray}
\textup{E}\left [ i_f(t) \right ] = 0, \nonumber \\
\textup{E}\left [ i_f(t)i_f(t+\theta) \right ] = 2\gamma_{G} (T)\delta \left (\theta \right).
\label{WNCorrelation}
\end{eqnarray}
Thus, the amplitude of the normalized correlator can be connected to the physical temperature
\begin{eqnarray}
\label{WNAmp}
\gamma_G (T)
& = & \frac{kT}{R_N}\frac{\omega_{po}}{I^2_c}
=\frac{2e}{\hbar}\frac{kT}{I_c} \beta_J= \nonumber \\
& = & \frac{kT}{E_{J_0}} 
= \frac{kT}{I_c R_N}\left( \frac{2e}{\hbar I_c C}\right)^{1/2}.
\label{WNAmpA}
\label{WNAmpB}
\end{eqnarray}
For instance, for a short JJ with a critical current $I_c = 0.1 \mu \textup{A}$ the noise amplitude, the normal resistance $R_N=100 \Omega$, $C = 20 pF$, one finds that $\gamma_G=10^{-3}$ corresponds to the temperature $T\sim 150 \textup{mK}$.
{\hg  As usual for numerical simulations in normalized units the reported quantities, as the Gaussian white noise amplitude, should be related to physical quantities through the system physical parameters: the critical current, the normal resistance, the capacitance, and the temperature of the Josephson junction. 
The correspondence between a temperature in the range of hundreds of $mK$ and the value $\gamma_G = 10^{-3}$ only highlights the connection to exemplify the meaning of the dimensionless quantity $\gamma_G$. }

The physical interpretation of the L\'evy process component, $\gamma_L$ in Eq.~(\ref{GLFincr}), can be reconducted to the L\'evy-Ito decomposition theorem \cite{Kyprianou06}. 
Shortly, this theorem states that L\'evy process is a mixture of a pure Poissonian-like jump process and a standard Brownian motion.
In the absence of the white noise, $\gamma_G=0$, the current fluctuations consist of finite jumps alternated with a standard Brownian erratic drift. 
%

\subsection{Current source and switching current measurements}
\label{sec:current}

In Eq.~(\ref{RCSJ}), the term $i_b$ represents an external bias current normalized to the critical current $I_c$. 
The essential point to retrieve information on the noise induced activation is to linearly ramp the external bias
\begin{equation}
 i_{b}(t)=i_0+\frac{di_b}{dt}t=i_0+v_b t ,
\label{BiasCurrent}
\end{equation}
where $i_0$ is an initial bias that can be reached with a very fast initial sweep (see the inset in Fig.~\ref{Figschematic}b).
Above $i_0$, the current is increased at the speed $v_b$.
According to Eq.~(\ref{grapheneWB}), when the bias increases, the potential $U$ is tilted, and the trapping energy barrier $\Delta U$ decreases (see Fig.~\ref{Figschematic}a) up to vanish for $i_b=1$.
During this process, fluctuations induced by the noise source $i_f$ eventually cause a passage over the top of the potential towards the running state, where a finite voltage $V =(\hbar /2e) \langle d\varphi /dt \rangle $ appears~\cite{Jos62,Jos74} (see Fig.~\ref{Figschematic}b). 
Schematically, the experimental procedure is as follows.
Above the crossover temperature~\cite{Gra84} $\TsCO=\hbar \omega_R / 2\pi k$, the quantum tunnel is negligible and the potential (\ref{grapheneWB}) is tilted at a rate $di_b/dt=v_b$ until the representative point of the system "runs down", i.e. it assumes a sizable speed that, in the Josephson counterpart, corresponds to a time phase variation producing a nonzero mean voltage across the junction (see Fig.~\ref{Figschematic}a).
At this point the corresponding bias value $i_b\equiv i_{SW}$ is recorded (see Fig.~\ref{Figschematic}b). 
Without noise this switching current reads $i_{SW}= 1$ (that is, the very meaning of the critical current $I = I_c$, the current at which the static state is unstable), while in the presence of fluctuations the switch occurs for $i_{SW,j} < 1$, because of the presence of a random term $i_f$.
However, repeating the process, i.e. ramping again the bias current from $i_b(0)=i_0$ the measured SC is different, for the random nature of $i_f$. 
This produces a distribution of the SCs, $i_{SW,j}$, where the index $j=1,N$ denotes the $j^{th}$ experiment (or realization).

{\hg The method of current tilting described in this Sect. acts on a special feature of graphene-based JJs, the potential (\ref{grapheneWB}), that is specific of graphene-based JJs. 
Also, the method of switching current analysis is particularly suitable, or perhaps strictly necessary, for nanodevices as those based on JJ, inasmuch the voltage corresponding to the critical current is a macroscopic signature of nanoscale features. 
Thus, the switching current analysis, carried out on graphene-based JJs characterized by the potential (\ref{grapheneWB}),  allows to investigate a nanoscale device through the analysis of a macroscopic, accessible quantity: the switching current.}

\begin{figure}[t!!]
\centering 
 \includegraphics[width=0.4\textwidth]{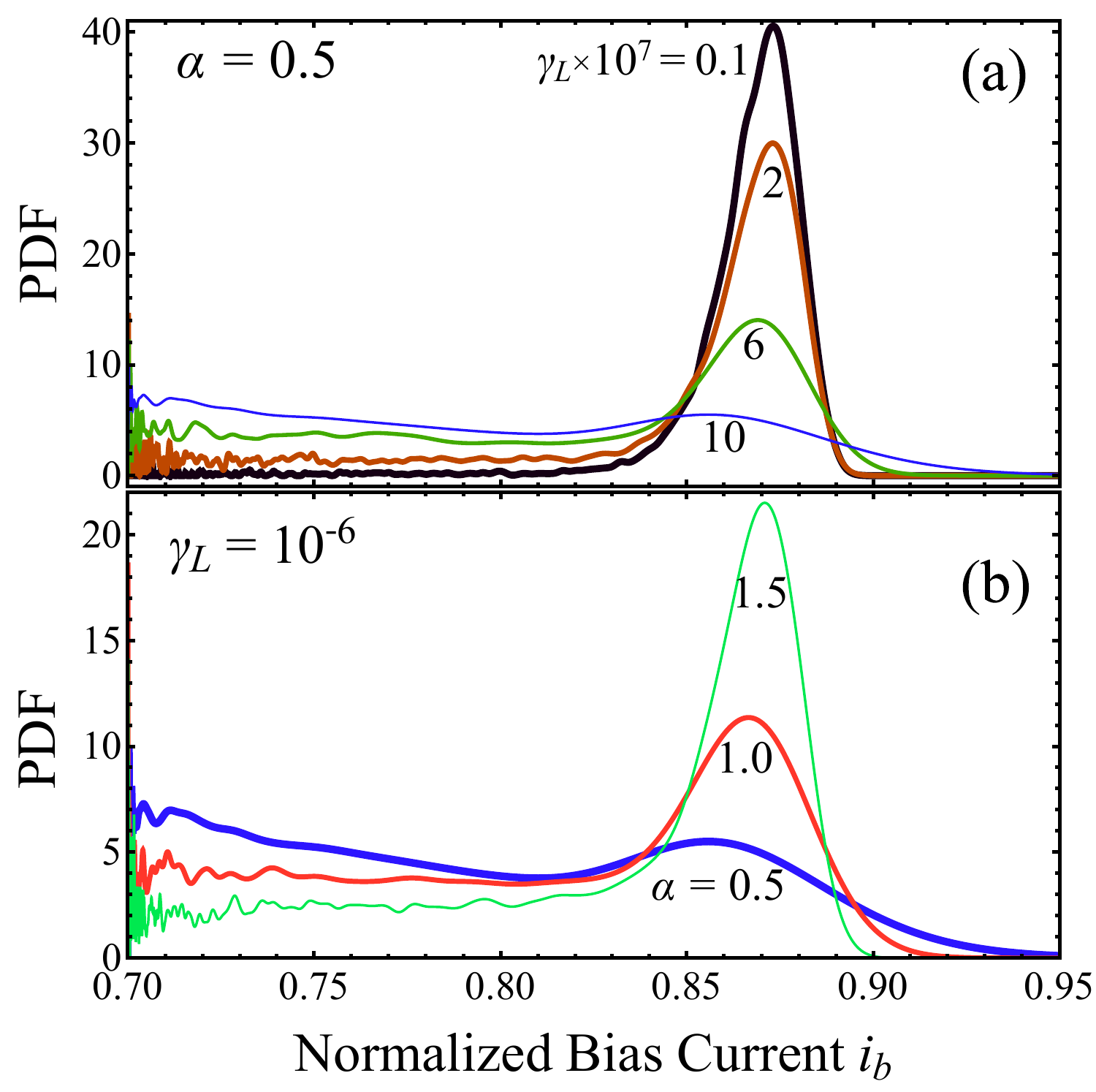} \\
\caption{(Color online) Switching current distributions for several L\'evy noise amplitudes (from the highest tail to to lowest) $\gamma_L=10^{-8}$, $2\cdot10^{-7}$, $6\cdot10^{-7}$, $10^{-6}$ and $\alpha=0.5$, and for different values of the L\'evy parameter (from the highest tail to the lowest) $\alpha=0.5$, $1.0$, $1.5$ and $\gamma_L=10^{-6}$, see panels a and b, respectively. 
The parameters of the system are: thermal noise amplitude $\gamma_G = 10^{-3}$, initial bias $i_0=0.7$, damping parameter $\bJ=0.10$, and ramping speed $v_b = 10^{-7}$. 
The current is ramped at a speed $v_b = 10^{-7}$.}
\label{Fig03}
\end{figure}

The distribution of the SCs is the quantity that we employ to characterize the noise features. 
That is, we analyze the SC data resulting from the above described procedure, and for instance shown in Fig.~\ref{Fig03}.
This analysis is the subject of the next Section.

\section{Switching Currents Analysis}
\label{sec:SCanalysis}

Let us suppose that the data are collected in the form of SC distributions as those of Fig.~\ref{Fig03}. Our objective is to determine the most effective methods to retrieve the noise features in graphene-based JJs. 

\subsection{Analysis of the escapes for the L\'evy noise case }
\label{sec:Levy}

In Fig.~\ref{Fig03} the SC distributions, computed in the presence of both Gaussian and L\'evy noise contributes, are shown for several values of the L\'evy noise amplitude and $\alpha$ L\'evy parameter, while the thermal noise amplitude is set to $\gamma_G=10^{-3}$. 
We observe that: 
\textit{i}) if $\alpha$ is fixed, as $\gamma_L$ grows the L\'evy noise effects are enhanced, so that the low-current tail of the SC distributions raises, while the peak decreases, see Fig.~\ref{Fig03}a; 
\textit{ii}) if the $\gamma_L$ value is fixed, as $\alpha$ increases the low-currents tail of the SC distributions is depleted and the peak grows (see Fig.~\ref{Fig03}b).
The position of the peak of the SC distributions is only slightly affected by the change of the L\'evy parameters, thus underlining the thermal origin of this spike.

Escapes over a barrier in the presence of L\'evy noise have been extensively investigated for the overdamped case~\cite{Dyb06,Dyb07,Dub08,Dub09,Che07}.
It has been found that the asymptotic behavior of the mean escape time under the effect of the L\'evy noise is of power-law type~\cite{Dub07,Che07}.
In fact, in the low noise intensity regime, the fat-tails property of the L\'evy noise allows for large outliers that dominate the escape process~\cite{Che07}.
Put it in other words, the fluctuations corresponding to the extreme values of the L\'evy distribution determine the majority of the switches and therefore, in this regime, the escape probability is independent of the energy barrier.
The assumption of a switching time $\tau$ independent of the bias current is consistent with the experimental results of Coskun \textit{et al}~\cite{Cos12} that the switching rate for low currents deviates from the expected exponential behavior and shows a plateau (see Fig.2d of Ref.~\cite{Cos12}). 
Accordingly, following \cite{Fulton74} the probability density function of the switching current, conditioned by the initial bias current $i_0$, can be assumed to be of the type
\begin{equation}
P(i_b|i_0) \propto \exp \left [ - {\mathcal K} i_b \right ],
\label{P_t}
\end{equation}
where ${\mathcal K }$ is a constant that depends upon the noise features.

\subsection{Analysis of the L\'evy noise properties through the moments}
\label{sec:moments}

\begin{figure}[t!!]
\centering \hspace{2.2cm}
 \includegraphics[width=0.5\textwidth]{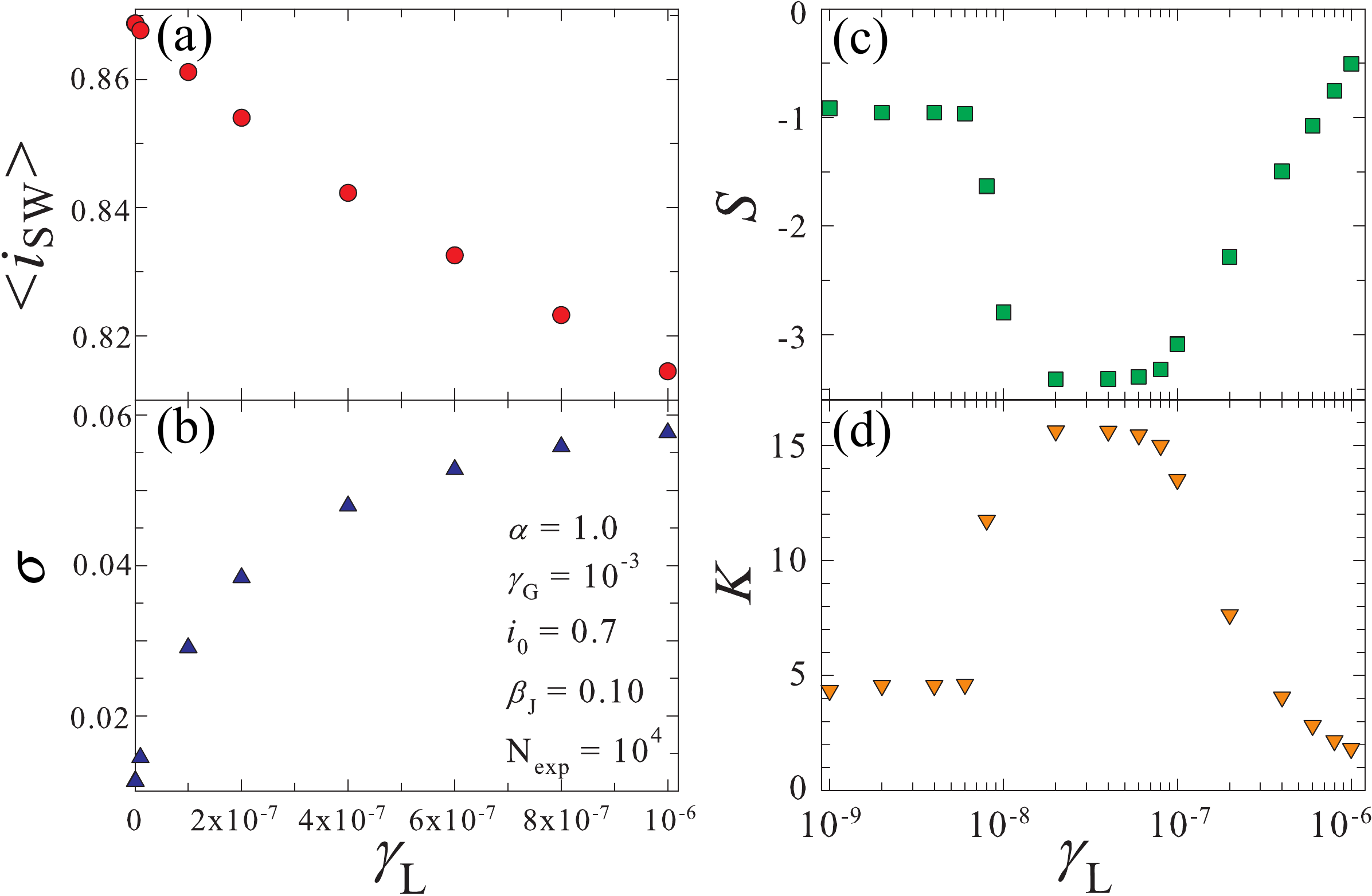} 
\caption{(Color online) Dependence of the moments upon the L\'evy noise intensity $\gamma_L$ for $\alpha=1.0$. We underline that the limiting values for negligible L\'evy noise amplitude of skewness and kurtosis agree with the theoretical estimates given in~\cite{Mur13}. 
The parameters of the system are: thermal noise amplitude $\gamma_G = 10^{-3}$, initial bias $i_0=0.7$, damping parameter $\bJ=0.10$, and ramping speed $v_b = 10^{-7}$.}
\label{Fig04}
\end{figure}

If the SC distributions are used to highlight the presence of L\'evy noise, a natural solution is to employ the moments of the SC distribution, as the skewness 
\begin{equation}\label{Skewness}
\tilde{S}[X]=\frac{\textup{E}\left [ \left ( X-\mu \right )^3 \right ]}{ \textup{E}\left [ \left ( X-\mu \right )^2 \right ]^{3/2}},
\end{equation}
where $X$ represents the random variable of the measurements, and $\mu$ is the expected value.
The quantity (\ref{Skewness}) can be estimated through the measured switching currents $i_{SW,j}$
\begin{equation}
\label{Skewness_observed}
S=\frac{\left \langle \left ( i_{SW,j}-\left \langle \iSW \right \rangle \right )^3 \right \rangle}{\sigma^3},
\end{equation}
here $\left \langle i_{SW} \right \rangle$ is the estimate of the mean switching current, and, as usual, $\sigma$ is the estimate of the standard deviation of the switching distribution.

Analogously, the fourth moment, the kurtosis, reads
\begin{equation}\label{Kurtosis}
\tilde{K}[X]=\frac{\textup{E}\left [ \left ( X-\mu \right )^4 \right ]}{ \textup{E}\left [ \left ( X-\mu \right )^2 \right ]^2},
\end{equation}
and the corresponding measured moment is
\begin{equation}
\label{Kurtosis_observed}
K=\frac{\left \langle \left ( i_{SW,j}-\left \langle \iSW \right \rangle \right )^4 \right \rangle}{\sigma^4}.
\end{equation}

If the SC of the JJ can be approximated by a linear system, Gaussian noise should result in a Gaussian distribution of the SC, therefore characterized by $S \simeq \tilde{S}[X] = 0$ and $K=3$. 
Unfortunately, even if the JJ were linear systems, the escape time threshold would introduce an intrinsic nonlinearity that deformates the SC distributions. 
In fact, the experimental~\cite{Mur13} and theoretical~\cite{Garg95} distributions show that $S=-1$ and $K=5$, thus underlining the nonlinear character of the escape measurements.
This has been confirmed in graphene-based junctions and ultrathin superconducting nanowires.
In fact Murphy \textit{et al.}~\cite{Mur13} have experimentally found for the SC distributions the values $S\approx -1$ and $K \approx 5$.

We go one step further, hypothesizing that deviations from the abovementioned values of the moments $S=-1$ and $K=5$ can be used to quantify the amplitude of the non-Gaussian noise input.
Numerical simulations of graphene based JJ indeed demonstrate that the moments depend upon the noise features, as we show in Fig.~\ref{Fig04}. 
In the Figure the intensity of thermal noise is kept fixed to $\gamma_G = 10^{-3}$, and the first four moments of the SC distributions show to be dependent on the L\'evy noise amplitude. 
In particular, it is evident that increasing the L\'evy noise amplitude $\gamma_L$, the mean switching current decreases (see Fig.~\ref{Fig04}a) and the standard deviation increases (see Fig.~\ref{Fig04}b), for the low currents tail of the SC distributions are enhanced (see Fig.~\ref{Fig03}). 
From the results shown in panels c and d of Fig.~\ref{Fig04} it is evident that the behavior of the higher moments is nonlinear, since the skewness shows a minimum and the kurtosis has a peak for the $\gamma_{L} \sim 2\cdot10^{-8}$. 
The method can be effectively used to determine the nature of the noise affecting the JJ, but has some limits when employed to measure the non-Gaussian component, i.e. to estimate the coefficient $\gamma_L$.
In fact, let us suppose that one wants to estimate the noise amplitude $\gamma_L$ through experimental measures of the skewness and kurtosis.
Because of the non-monotonic behaviors shown in Fig.~\ref{Fig04}c,d, in close proximity of the minimum (maximum) of the skewness (kurtosis) distribution, a single value of $\gamma_L$ cannot be univocally determined by the measurement of $S$ ($K$). Furthermore, also in the amplitude regions in which the moment's behavior tends to become flat, the estimation of $\gamma_L$ from the knowledge of $S$ and $K$ is not easy, and thus the accuracy of the method becomes poor in this region.
This is not surprising, for it is known that the parameter estimate is better performed with statistical tests rather than through the evaluation of the moments.
(Such an analysis is the subject of the next Sect.~\ref{Analysis}.)

\begin{figure}[t!!]
\centering 
\includegraphics[height=7cm]{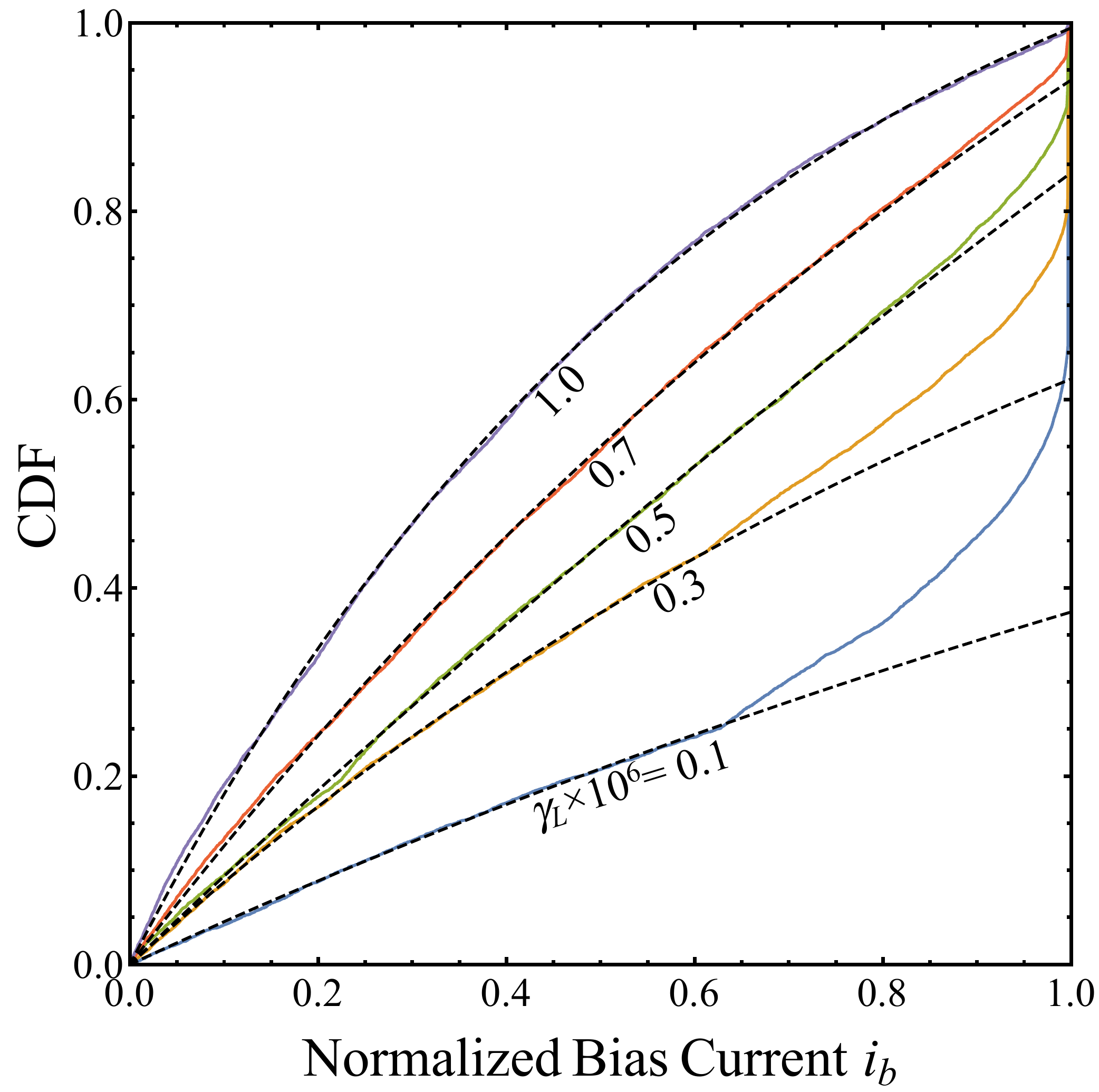} 
\caption{(Color online) 
Cumulative SC distributions when only a L\'evy noise source, with $\alpha = 1$ and $0.1 \leq \gamma_L \times 10^6 \leq 1.0$ (the arrow indicates the order) is considered, i.e. $\gamma_G=0$. The parameters of the system are: initial bias $i_0=0.0$, damping parameter $\bJ=0.10$, and ramping speed $v_b = 10^{-7}$.
The solid line represents numerical simulations of Eq.(\ref{RCSJ}), while the dotted line is the interpolation of the exponential behavior of Eq.~(\ref{P_t}).
}
\label{Fig05}
\end{figure}
%

\subsection{Analysis of the L\'evy noise properties through the switching current distributions}
\label{Analysis}

If JJs are used to reveal the presence of L\'evy noise in graphene based junctions, the starting point is to analyze the response to an unknown random perturbations.
We first notice that the L\'evy noise alone, as shown in Fig.~\ref{Fig05}, produces an exponential dependence in the SC cumulative distribution.
This is the counterpart of the PDF behavior at low current in Fig.~\ref{Fig03}. 
In fact the curves obtained by interpolation of the data with an exponential model, the dotted lines of Fig. \ref{Fig05}, are in very good agreement with numerical simulations (solid lines) untill the assumption of Eq.(\ref{P_t}) holds, that is in the fat tail region of the PDFs of Fig. \ref{Fig03}. 
For low bias currents, the exponential behavior is due to the heavy tail character of L\'evy noise.
For high values of the noise intensity $\gamma_L$, the tails of the L\'evy distribution dominates the switching dynamics, resulting in the early passages of the JJ to a finite voltage.
Instead, when the L\'evy noise intensity is very low, i.e. $\gamma_L=10^{-7}$ in Fig.~\ref{Fig05}, the switches induced by L\'evy flights occur at a lower rate.
Thus, only in correspondence of high potential slopes, i.e. $i_b\simeq 0.65$ for $\gamma_L=10^{-7}$ in Fig.~\ref{Fig05}, the peaked behavior of the L\'evy noise distribution in the neighbourhood of zero induces a steep rise of the cumulative distribution function (CDF). 

\begin{figure}[t!!]
\centering
(a) \\
\includegraphics[height=7cm]{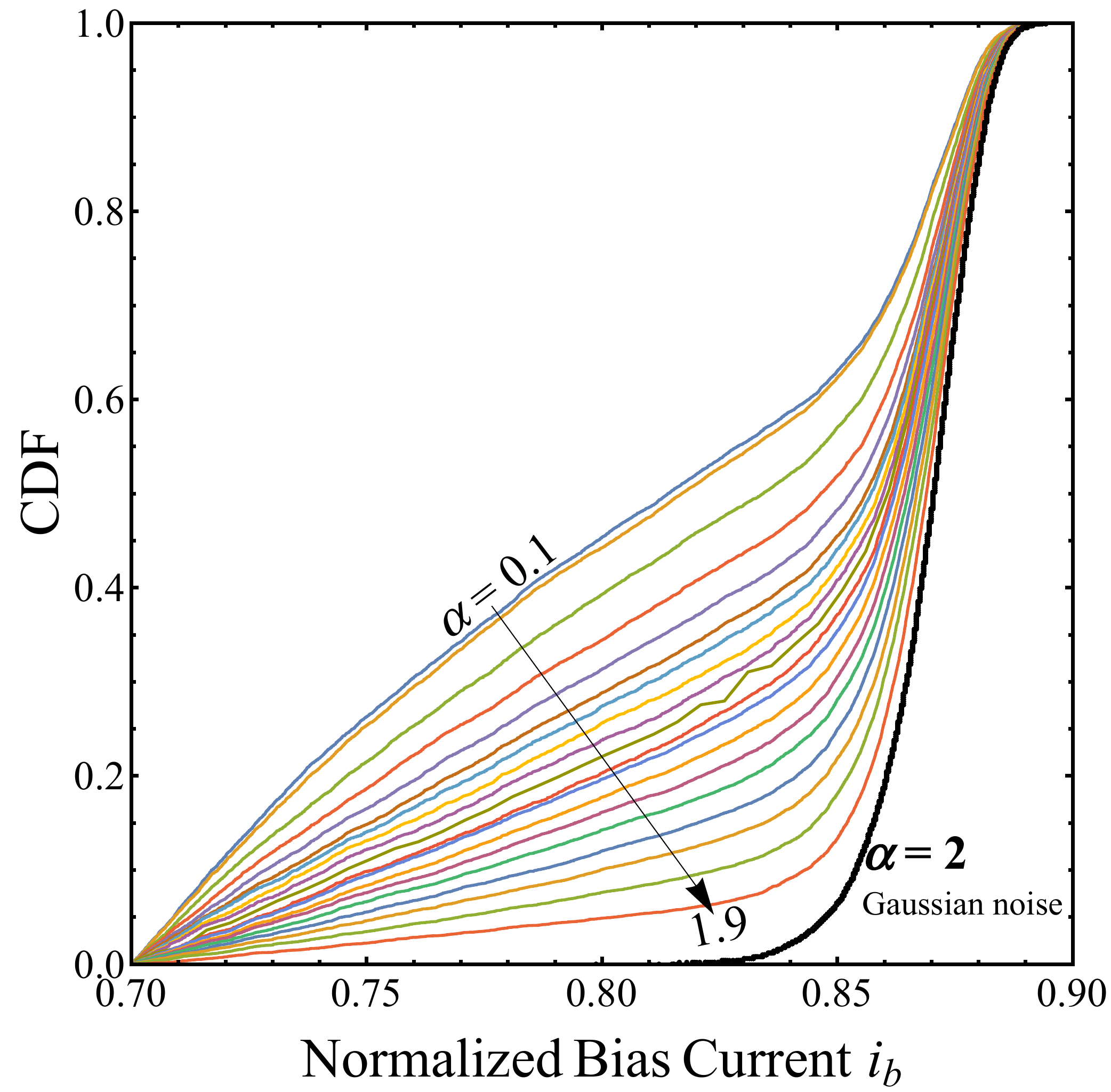} \\
(b) \\
\includegraphics[height=7cm]{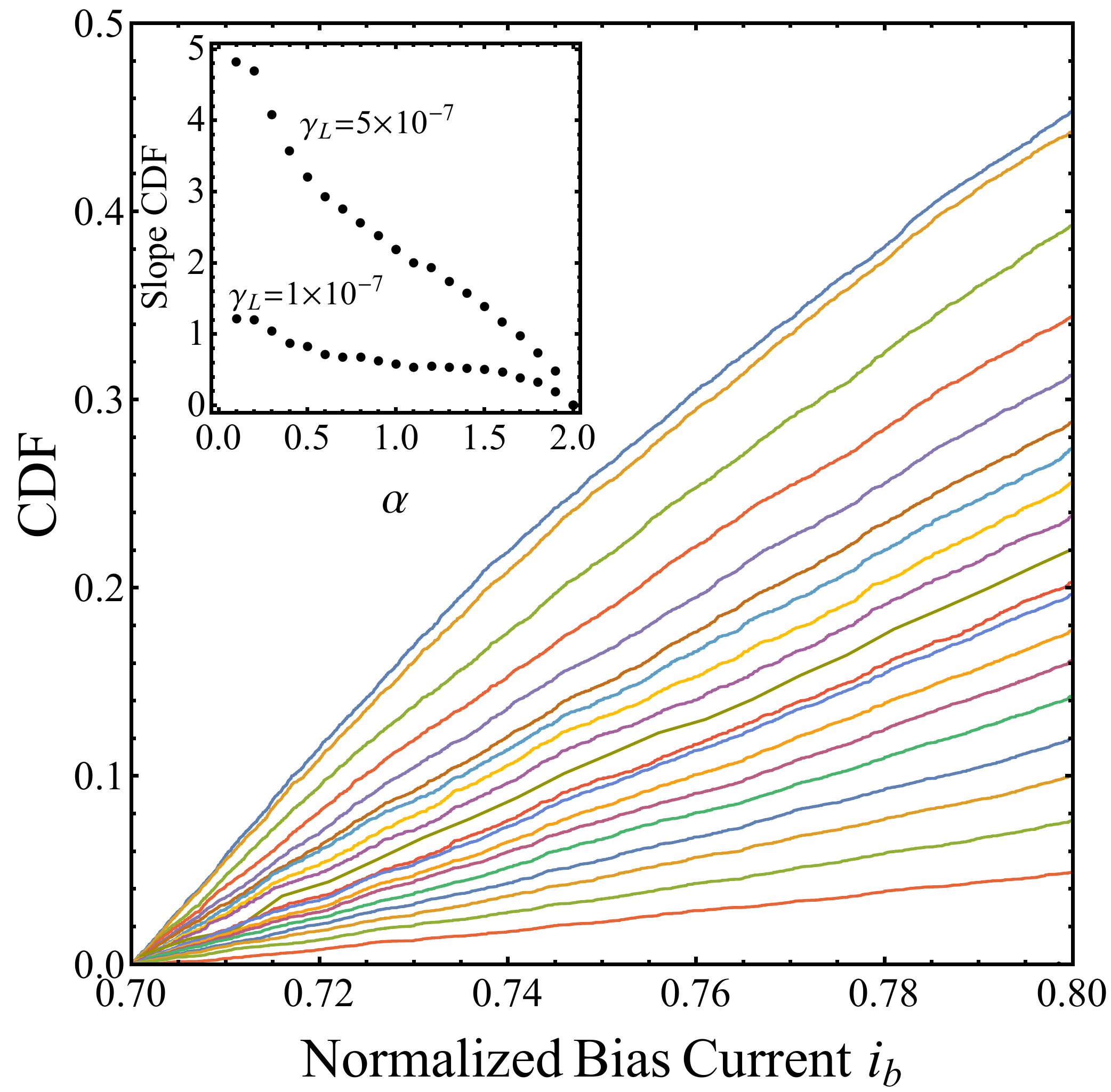}
\caption{(Color online) (a) CDFs in the presence of both Gaussian and L\'evy noise sources with amplitude $\gamma_G=10^{-3}$ and $\gamma_L=5\cdot10^{-7}$, respectively, and for $\alpha\in(0\div2]$ with step $0.1$ in the order indicated by the arrow. The Gaussian noise intensity is such that it matches the energy activation for bias $i_b \simeq 0.87$. The other parameters of the system are: initial bias $i_0=0.7$, damping parameter $\bJ=0.10$, and ramping speed $v_b = 10^{-7}$. (b) Enlargement of the initial linear part of results in panel a. In the inset, the slope $b$ of the cumulative distribution of the SC as a function of the L\'evy distribution parameter $\alpha$ for $\gamma_L=(5\cdot10^{-7},10^{-7})$.}
\label{Fig06}
\end{figure}

The detection problem of the L\'evy noise presence amounts to retrieve the abovementioned exponential behavior and the values of the parameters $\alpha$ or $\gamma_L$.
Having established that L\'evy noise gives rise to an exponential behavior, it remains to be devised an effective method to determine the noise properties from the SC.

{\hg In doing so, one should first assume that the unavoidable random term, statistically associated to thermal noise, is described by Gaussian distributed fluctuations \cite{Bar82}. 
If one ascertains the plausibility of some further non-Gaussian disturbances, it is necessary to formulate, on physical grounds, a statistical model for the additional random term.
The latter model is the basis to retrieve, with respect to the pure Gaussian hypothesis, the statistical plausibility of the presence of L\'evy flights. 
Thus, if the hypothesis of pure Gaussian noise is rejected on a statistical basis and the L\'evy flight hypothesis is acceptable on physical grounds, we propose to retrieve the $\alpha$-stable parameter of the L\'evy noise from the analysis of  the switching current distribution.

To describe the abovementioned procedure in concrete terms, let us }
begin with simulated SC in the presence of different sources of input noise.
The results are collected in Fig.~\ref{Fig06} in the form of  cumulative distribution function; the two panels display the percentage of switching from the superconductive to the normal state that have occurred prior to a certain value of the bias current $i_b$ while repeatedly ramping the current, as per Eq.~(\ref{BiasCurrent}).
Figure~\ref{Fig06} also demonstrates a qualitative finding: the L\'evy component gives a clear initial linear slope, that is made evident in the enlargement of Fig.~\ref{Fig06}b.
The change with respect to the pure Gaussian noise case is a very effective distinctive feature that can be exploited to decide if together with the Gaussian noise there is a L\'evy statistics component.
In fact, even with very few data, a linear CDF can be effectively distinguished from the zero background for low currents, resulting when only the Gaussian noise contribution is considered.

To detect the $\alpha$ value we show the association between the CDF slope and the L\'evy flight parameter $\alpha$ in the inset of Fig.~\ref{Fig06}b. 
From this Figure it is evident that the slope of the cumulative distribution depends upon the parameter $\alpha$: the smaller $\alpha$ the greater the slope.
This is not surprising, because the case $\alpha = 2$ corresponds to the Gaussian case, where a neat peak without tails is expected.
The change in the slope offers the possibility to determine the parameter $\alpha$.

To make the analysis quantitative, we employ the Kolmogorov-Smirnov (KS) \cite{Numrecip} test to distinguish the different outcomes. 
We have performed the KS test on the data of Fig.~\ref{Fig06}, comparing the pure Gaussian case ($\gamma_G=10^{-3}$) with the mixture of the same Gaussian source and an additional L\'evy noise ($\alpha = 1.9$) of amplitude $\gamma_L = 5 \cdot 10^{-7}$.
The result is that the test is very effective, to say the least, as the KS statistics reads $D=0.072$.
For a sample size of $10^4$ switching currents, it corresponds to a $p$-value of $10^{-23}$. 
It is interesting to note that this very high significance is due to a qualitative change between the L\'evy flight and the Gaussian noise.
In fact, if one compares two L\'evy-noise type, say $\alpha=1.3$ with $\alpha = 1.4$ (see Fig.~\ref{Fig06}), the KS statistics reads $D=0.0257$, and the corresponding $p$-value ramps up to $p= 0.0026$, about $20$ orders of magnitude above.
One concludes that the analysis of the SC is particularly effective in detecting the presence of noise of the L\'evy type.
The KS test also allows to find the minimum number of the measurements to detect the presence of a L\'evy noise component (of a given amplitude) with a prescribed $p$-value, or conversely the $p$-value for a given number $N$ of experiments.
The results are shown in Table \ref{table_N}, for two values of the L\'evy noise amplitude $\gamma_L$.
The data demonstrate that with a number of experiments in the order of $10^3$ it is possible to achieve a $p$-value below $1\%$.
As expected, the lower the noise level, the higher the number of data necessary to confirm the presence of the L\'evy noise component. 
Thus, Table \ref{table_N} can also be used to rule out, for a given confidence level $p$, the presence of noise at the amplitude reported.

\begin{table}[]
\centering
\caption{The $p$ values as a function of the number of data for two values of the L\'evy noise amplitude $\gamma_L$.
The Gaussian noise amplitude is $\gamma_g=10^{-3}$, the L\'evy noise parameter is $\alpha = 1.9$. }
\label{table_N}
\begin{tabular}{|c|c|c|}
\hline
$\gamma_L$ & $N$ & $p$ \\ \hline
\multirow{4}{*}{$10^{-7}$} & $10^2$ & 0.906 \\ \cline{2-3} 
 & $5\times 10^2$ & 0.665 \\ \cline{2-3} 
 & $10^3$ & 0.062 \\ \cline{2-3} 
 & $5\times 10^3$ & 0.004 \\ \hline
\multirow{4}{*}{$5\times10^{-7}$} & $10^2$ & 0.813 \\ \cline{2-3} 
 & $5\times 10^2$ & 0.111 \\ \cline{2-3} 
 & $10^3$ & 0.001 \\ \cline{2-3} 
 & $5\times 10^3$ & $3\times 10^{-12}$ \\ \hline
\end{tabular}
\end{table}

We conclude that the switching currents are potentially interesting to reveal, or to exclude, the presence of an $\alpha$-stable, L\'evy noise type, as the switchings are very sensitive to the presence of heavy tails noise disturbance.

\section{Conclusions}
\label{Conclusions}
We have addressed the problem to detect the presence of non-Gaussian noise in Josephson systems through the analysis of the distribution of the switching currents. 
The question is of relevance in the context of material analysis, specifically for graphene-based JJ where there is the indication that nonequilibrium noise, induced for instance by a laser beam or by anisotropically distributed atoms, can have an infinite, L\'evy type, variance.
More technically, an infinite variance entails a finite probability that a fluctuation passes any given finite threshold, however large, as the probability of large excursions only decays with an exponent $-\alpha$.
We propose a method for analyzing the switching current distributions of JJs subject to an unknown source of noise.
The method offers some distinct advantages when the noise is characterized by fat tails, i.e. by a finite probability of an infinite fluctuation.
This type of noise usually poses a serious difficulty to the experimentalist, for it requires extremely long times to reconstruct the behavior at large values.
Thus, to determine the value of the parameter $\alpha$ demands for long experiments (or simulations) to explore extreme values.
In contrast, sweeping the bias is very effective, because the bias increase lowers the trapping energy barrier, and therefore in a given ramp time the energy barrier vanishes and a switch event is recorded. 
The probability of a particle to overcome a barrier when subject to L\'evy noise is independent of the barrier height, at least when the ratio between the noise intensity and the energy barrier is high enough~\cite{Che07}.
This is remarkably different from the Gaussian noise case, where the probability to overcome the barrier depends exponentially on the barrier energy.
Also, if both components contribute to the overall noise level, they do not interfere, because they produce switching at different bias levels: the L\'evy noise in the lower part of the distribution, the Gaussian noise when the energy barrier becomes comparable to the noise energy.
The practical consequence for the analysis of graphene-based JJ is that the study of the switching current distribution is very effective in revealing the presence of L\'evy noise.
The analysis of the SC moments can be performed to estimate the amplitude of the L\'evy noise.
It is however more efficient to employ a statistical test, such as the Kolmogorov-Smirnov test, that can lead to an upper bound for the L\'evy noise level.
For instance, if the number of the measured switching currents is in the order of the thousands, it suffices to rule out at a confidence level of $1\%$ that the amplitude of the L\'evy noise is $10^{-7}$, when the Gaussian noise amplitude is $10^{-3}$, that corresponds, in the normalized units we are using, to about $150 \, mK$.

The general conclusion is that the SC distributions in the cases of L\'evy and Gaussian noise are remarkably different, and the two cases can be statistically distinguished even if few data are available.
{\hg We stress that the use of the graphene-based model, especially Eq.(\ref{grapheneWB}), is prompted by the observation that L\'evy flight noise has been postulated only in the specific case of graphene-based JJ \cite{Cos12,Bri14,Gat16}, and that to reveal its presence might be of particular relevance for material issues that only pertain graphene.}

Apart material issues for the detection of intrinsic L\'evy noise, the efficiency of the method paves the way towards potential applications for the detection of extrinsic noise.
We speculate that it could be possible to develop bolometers based on graphene-based Josephson detectors, once the system has been adequately calibrated. 
Such calibration, however, requires a careful analysis of the parameter space, that is, in our view, the priority for the research in this direction.

\vspace{-0.3cm}

\section*{Acknowledgements}
\vspace{-0.5cm}
C.G. has received funding from the European Union FP7/2007-2013 under REA grant agreement no 630925 -- COHEAT and from MIUR-FIRB2013 -- Project Coca (Grant No.~RBFR1379UX). 
V.P. acknowledges INFN, Sezione di Napoli (Italy) for partial financial support.

\bibliographystyle{apsrev4-1}
\bibliography{biblio}

\end{document}